\documentclass[epj]{webofc}
\usepackage[utf8]{inputenc}
\usepackage[english]{babel}
\usepackage{amsmath, amssymb,graphicx}
\usepackage{caption}
\usepackage{subcaption}
\pdfoutput=1 
\usepackage{amsfonts}
\usepackage{ulem}
\usepackage[varg]{txfonts}

\newcommand{\nn}{\nonumber}
\newcommand{\be}{\begin{equation}}
\newcommand{\ee}{\end{equation}}
\newcommand{\bea}{\begin{eqnarray}}
\newcommand{\eea}{\end{eqnarray}}

\newcommand{\e}{\mathrm{e}}

\newcommand{\Tr}{\text{Tr}}

\newcommand{\dd}{\partial}

\newcommand{\tG}{\tilde{G}}
\newcommand{\tS}{\tilde{\Sigma}}

\newcommand{\tK}{\tilde{K}}

\newcommand{\mF}{\mathcal{F}}

\woctitle{Quarks 2016}

\begin{document}
\title{On $1/N$ diagrammatics in the SYK model\\ beyond the conformal limit}
\author{\firstname{Irina} \lastname{Aref'eva}\inst{1}\fnsep\thanks{\email{arefeva@mi.ras.ru}} \and
        \firstname{Mikhail} \lastname{Khramtsov}\inst{1}\fnsep\thanks{\email{khramtsov@mi.ras.ru}} \and
        \firstname{Maria} \lastname{Tikhanovskaya}\inst{1}\fnsep\thanks{\email{tikhanovskaya@mi.ras.ru}}}

\institute{Steklov Mathematical Institute, Russian Academy of Sciences,\\Gubkina str. 8, 119991, Moscow, Russia      }

\abstract{In the present work we discuss aspects of the $1/N$ expansion in the SYK model, formulated in terms of the semiclassical expansion of the bilocal field path integral. We derive cutting rules, which are applicable for all planar vertices in the bilocal field diagrams. We show that these cutting rules lead to novel identities on higher-point correlators, which could be used to constrain their form beyond the solvable conformal limit. We also demonstrate how the cutting rules can simplify the computation of amplitudes on an example of the six-point function.
}
\maketitle
\section{Introduction and summary of results}
The SYK model \cite{Sachdev92,Kitaev,MScomments,Kitaev17} is a maximally chaotic  quantum mechanical many-body model that is solvable at large $N$. It also displays emergent conformal symmetry in the strong coupling limit, that is broken both spontaneously and explicitly. Maximal chaos and emergent conformal symmetry at strong coupling motivate using the SYK model as a holographic toy model for quantum near-extremal black holes \cite{Kitaev,Maldacena16,Sachdev15}. 

The SYK is a model of $N$ interacting Majorana fermions, living in $(0+1)$ (Euclidean) dimensions. The Hamiltonian is given by \cite{Kitaev}:
\be
H = \frac{1}{4!} \sum_{i j k l} j_{ijkl} \chi_i \chi_j \chi_k \chi_l\, \label{Hsyk}
\ee
The couplings $j_{ijkl}$ are randomized with respect to the Gaussian distribution $P(j_{ijkl}) = \sqrt{\frac{N^3}{12\pi J^2}}
\ \e^{-\frac{N^3 j_{ijkl}^2}{12 J^2}}$. 
The correlation functions and other physical quantities in the model are supposed to be disorder-averaged, using the rules
\be
\langle j_{ijkl} \rangle = 0\,, \qquad  \langle j_{ijkl}^2 \rangle = \frac{3! J^2}{N^3} \label{disorder2point}
\ee
At finite temperature the constant $\beta J$ admits the role of the dimensionless coupling; in the IR limit the model is in the strong coupling regime $\beta J \gg 1$. 
At large $N$ limit in the language of Feynman diagrams in terms of Majorana fermions all correlation functions of the model are dominated by the melonic graphs and their cuts \cite{Kitaev,Gross17-3pt,Gross17}. 

An alternative formulation of the model can be obtained by averaging over the disorder in the partition function, introducing colorless bilocal auxiliary fields $\tG(\tau_1, \tau_2)$ and $\tS(\tau_1, \tau_2)$ and integrating out the fermions (see \cite{Kitaev17} for detailed derivation):
\be
Z= \int D\tG D \tS \ \e^{-S[\tG, \tS]}\,; \label{path-int}
\ee
where the action in terms of $G$, $\Sigma$ fields has the form:
\be\label{action}
S[\tG, \tS] = -\frac{N}{2} \Tr \log(\dd_\tau - \hat{\tS}) +\frac{N}{2}\int d\tau d\tau' \left(\tS(\tau, \tau') \tG(\tau, \tau') - \frac{J^2}{4} \tG(\tau, \tau')^4\right)\,,
\ee
(here and henceforth the notation $\hat{F}$ means the integral operator with the kernel $F(\tau, \tau')$). This path integral describes accurately describes the first few orders in $1/N$, where the replica-diagonal assumption is valid\footnote{The replica-nondiagonal saddle points of the disorder-averaged free energy are discussed in \cite{AKTV,AKTV-Quarks,AV}.}. The factor $N$ in the action justifies the usage of the steepest descent method to evaluate the path integral at large $N$. Thus the large $N$ expansion of SYK in terms of the bilocal fields is a semiclassical expansion for the path integral (\ref{path-int}). The fermionic correlators of the SYK model are expressed in terms of the correlators of the field $\tG$: 
\be
\langle \chi_{i_1}(\tau_1) \chi_{i_2}(\tau_2) \dots \chi_{i_{2n-1}}(\tau_{2n-1}) \chi_{i_{2n}}(\tau_{2n}) \rangle = \langle \tG(\tau_1, \tau_2) \dots \tG(\tau_{2n-1}, \tau_{2n}) \rangle\,.
\ee
In the strong coupling limit $\dd_\tau \to 0$ the action (\ref{action}) is conformally invariant. The conformal parts of all correlation functions in SYK can be computed analytically \cite{Kitaev,MScomments,Gross17}, however on the level of $4-$ and higher-point functions there are contributions that break the conformal symmetry and dominate in the $\beta J \to \infty$ limit. The non-conformal contributions come from the reparametrization soft mode \cite{MScomments,Polchinski16,Bagrets16,Kitaev17,Mertens17}, which is dual to the gravitational mode in the bulk theory. Because of this, the studies of the non-conformal corrections to the correlation functions in SYK can be useful to gain insight into the full bulk dual of SYK. 

In this work we discuss the technical tools which could be used to perform such calculations. Our main result are the cutting rules for planar diagrams. These cutting rules are exact in $\beta J$, and can be applied at any order of the $1/N$ expansion to either obtain additional constraints on the higher-point correlators beyond the conformal limit, or to simplify the calculations, as we demonstrate on the example of the non-conformal six-point function. 

\section{Systematic $1/N$ expansion in SYK}

Suppose $G$, $\Sigma$ is the large $N$ saddle point of the path integral (\ref{path-int}). We define new variables
\be
g = |G|(\tG - G)\,, \quad \sigma = \frac{1}{J^2 |G|} (\tS - \Sigma)\,,
\ee
and perform the semiclassical expansion in the bilocal fluctuation fields $g$, $\sigma$.
The action (\ref{action}) in terms of these fluctuation fields reads 
\bea
S[g, {\sigma}]&=& S_0[G, \Sigma]-\frac{N}{2} \Tr \log(1-J^2  \hat{G} \circ \widehat{\sigma |G|}) \nn\\
&+&\frac{N J^2}{2} \int d\tau d\tau' \left(\sigma g+|G|G\sigma - \frac14 g^4{1\over |G|^4} - g^3{G\over |G|^3}-\frac32 g^2{G^2\over |G|^2}\right)\,. \label{action-fluctuations}
\eea
We consider the diagrammatics for this action. Every propagator of a bilocal field carries a factor of $1/N$, whereas every vertex carries a factor of $N$. Therefore, the degree of a bilocal graph, contributing to a correlator, in the $1/N$ expansion is given by 
\be
\deg = L - V
\ee
where $L$ is the amount of lines in the diagram and $V$ is the number of vertices.

\subsection{Propagators}

The two-point functions of the bilocal fields are determined by the quadratic part of the action (\ref{action-fluctuations}) \cite{MScomments}:
\bea
&& S^{(2)} = -\frac{N J^2}{12} \int d^4\tau\ \sigma(\tau_1,\tau_2)\tilde{K}(\tau_1, \tau_2, \tau_3,\tau_4)\sigma(\tau_3,\tau_4) \nn\\&&
+\frac{N J^2}{2}\int d\tau_1 d\tau_2\left[ g(\tau_1,\tau_2)\sigma(\tau_1,\tau_2) - \frac32 g(\tau_1,\tau_2)^2\right]\,.\label{S2}
\eea
Here $\tK$ is the symmetric ladder kernel \cite{MScomments}:
\be \tilde{K}(\tau_1, \tau_2, \tau_3, \tau_4) =-3 J^2 |G(\tau_1, \tau_2)| G(\tau_1, \tau_3) G(\tau_2, \tau_4) |G(\tau_1, \tau_4)| \label{kernel}\ee
Inverting the quadratic form, we find the propagators:  
\bea
&& \langle g(\tau_1, \tau_2)  g(\tau_3, \tau_4) \rangle = \frac{2}{3N J^2} \left(\frac{\tK}{1-\tK}\right)(\tau_1, \tau_2; \tau_3, \tau_4)\,; \label{gg}\\
&& \langle g(\tau_1, \tau_2)  \sigma(\tau_3, \tau_4) \rangle = \frac{2}{N J^2}  \left(\frac{1}{\tilde{K}-1}\right)(\tau_1, \tau_2; \tau_3, \tau_4)\,; \label{gs}\\
&& \langle \sigma(\tau_1, \tau_2)  \sigma(\tau_3, \tau_4) \rangle = \frac{6}{N J^2}  \left(\frac{1}{\tilde{K}-1}\right)(\tau_1, \tau_2; \tau_3, \tau_4)\,. \label{ss}
\eea
The geometric series of the inverse ladder kernel in (\ref{gg}) is precisely the sum of the ladder diagrams, contributing to the fermionic four-point function. It can be computed explicitly by diagonalizing the ladder kernel in the conformal limit \cite{Kitaev,MScomments,Polchinski16}. Schematically, one obtains 
\be
\langle g\ g \rangle = \mF_{\text{non-conformal}} + \mF_{\text{conformal}},.
\ee
\begin{itemize}
\item Conformal part: $\mF_{\text{conformal}} \sim \sum_{h_m} c_{\Delta, m} F_{\Delta, m}(1,2,3,4)$- has the form of the conformal partial wave expansion in a $1$D CFT.
\item Non-conformal part: \be\mF_{\text{non-conformal}} \sim \beta J \sum_n \text{Res}_{h=2} \left[\frac{k(h,n)}{1-k(h,n)} \rho(h) \right] \psi_{2,n}^*(\tau_1, \tau_2) \psi_{2, n}(\tau_3, \tau_4)\,.\label{gg-nonconformal}\ee
Here $k(h, n)$ is the ladder kernel eigenvalue, and $\psi_{2,n}$ are the $h=2$ eigenfunctions of the symmetric ladder kernel (\ref{kernel}).
\end{itemize}
The non-conformal contributions to two-point functions (\ref{gs}),(\ref{ss}) are constructed analogously. 

\subsection{Vertices}

The action (\ref{action-fluctuations}) contains two types of vertices. We have two vertices for the field $g$ and an infinite series of vertices from the field $\sigma$ coming from the logarithm. One can show \cite{AKT} that the $g$-vertices correspond to the fermionic graphs labeled as contact diagrams by Gross and Rosenhaus \cite{Gross17-3pt,Gross17}. Meanwhile, the $\sigma$-vertices correspond to the planar diagrams. Based on this, we will refer to $g$-vertices as the \textit{contact} vertices, and to $\sigma$-vertices as the \textit{planar} vertices. 

\begin{figure}
\begin{center}
\includegraphics[scale=.50]{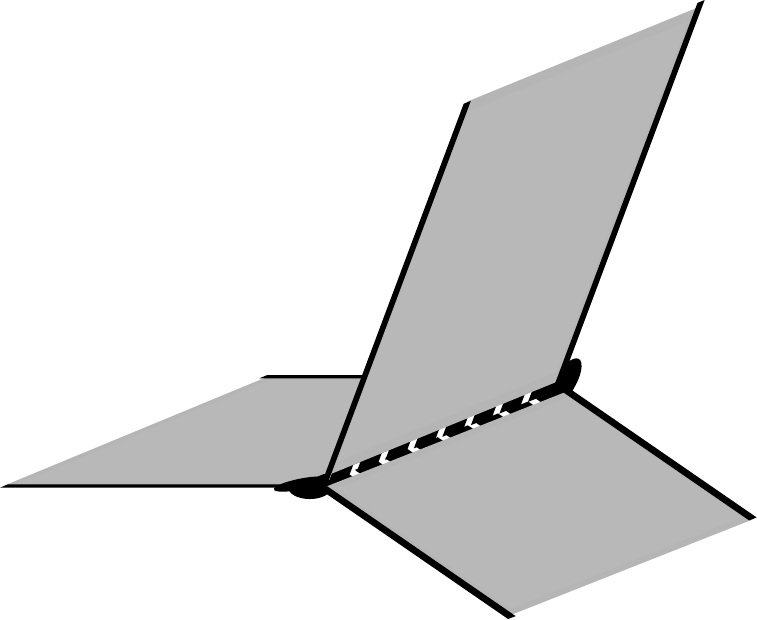}A.\quad 
\includegraphics[scale=.44]{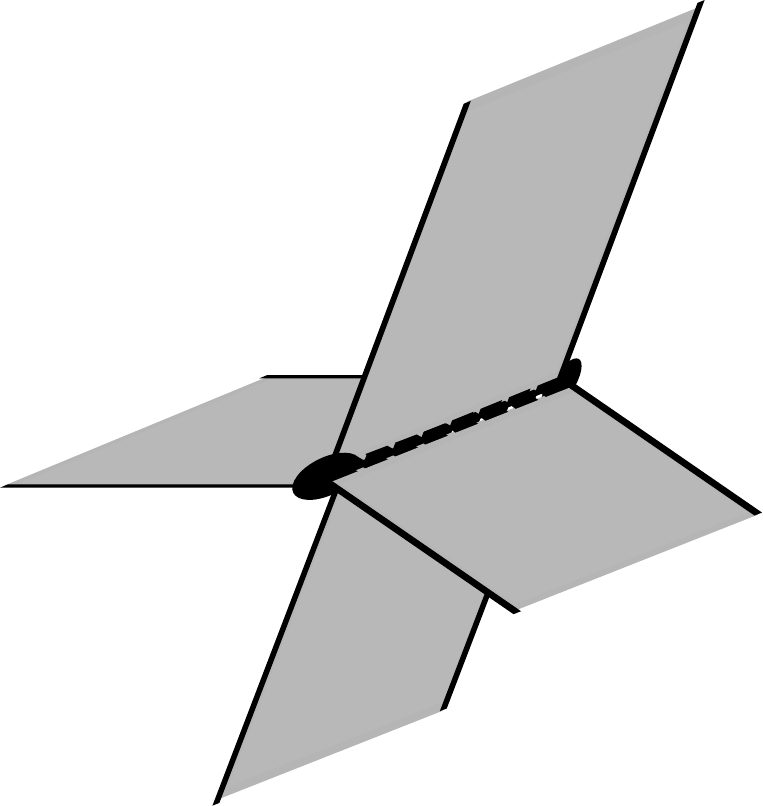}B. 
\caption{$3$-point and $4$-point contact vertices. }
\label{Fig:contact}
\end{center}
\end{figure}

\paragraph{Contact vertices}
From the action (\ref{action-fluctuations}) we obtain two contact vertices: 
\bea V_{g}^{3} = -\frac{N J^2}{2\ G(\tau_1, \tau_2) |G(\tau_1, \tau_2)|} g(\tau_1, \tau_2)^3\,; \qquad V_{g}^{4} = -\frac{N J^2}{8\ G(\tau_1, \tau_2)^4} g(\tau_1, \tau_2)^4\,.\eea
They are drawn on the Fig.\ref{Fig:contact}. 

\begin{figure}[t]
\begin{center}
\includegraphics[scale=.4]{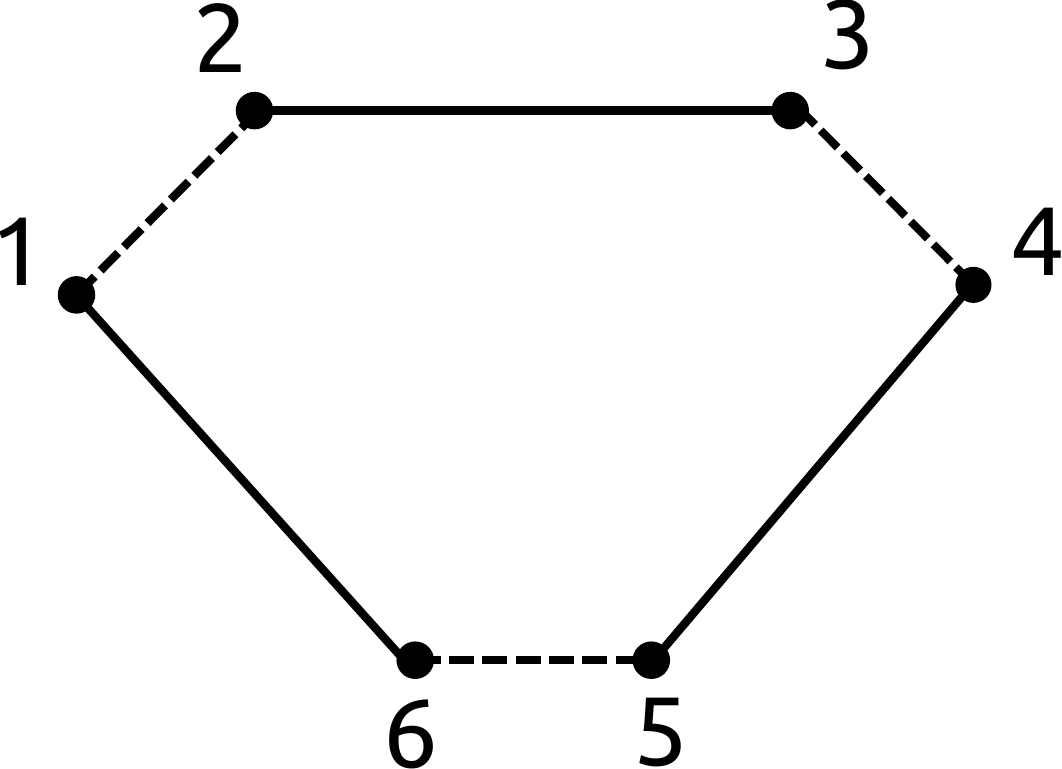}A.\qquad
\includegraphics[scale=.3]{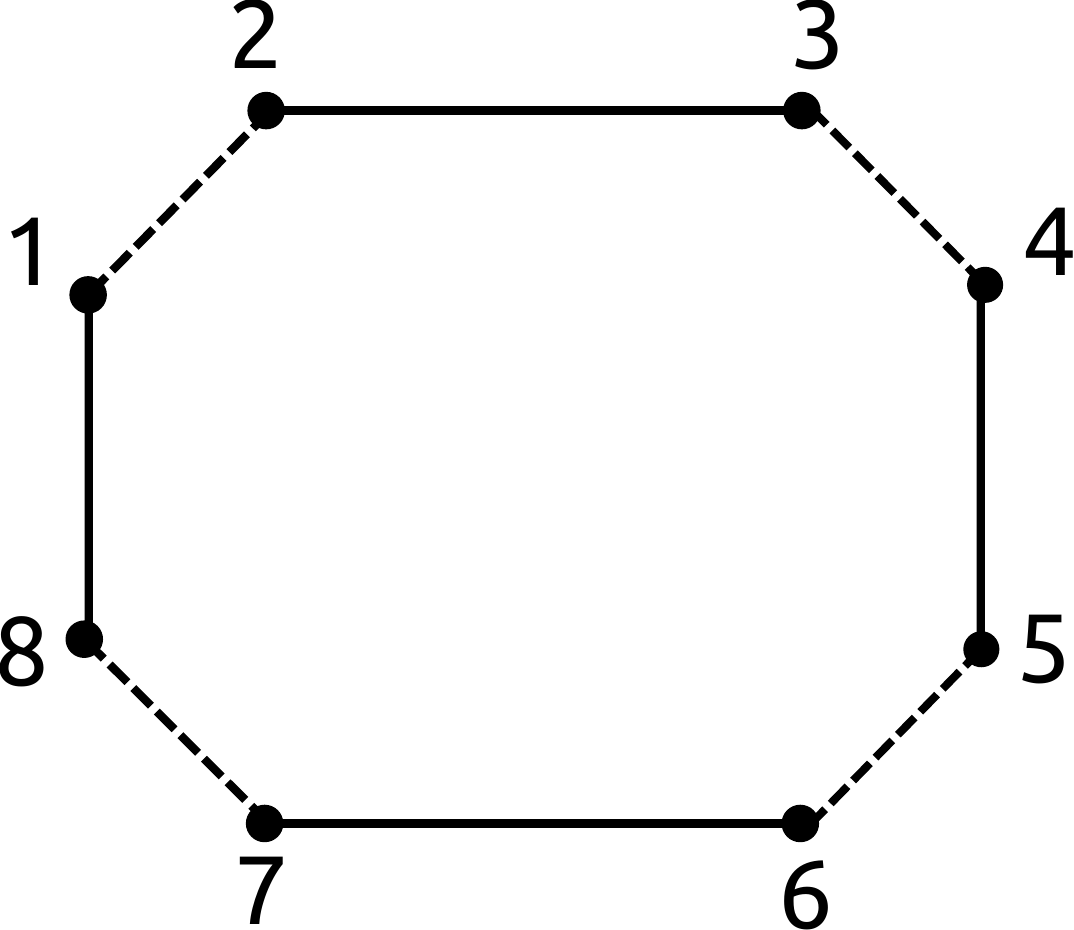}B.\\
\caption{\textbf{A}: Amputated irreducible planar $3$-point vertex. \textbf{B}: Amputated irreducible planar $4$-point vertex}
\label{diagrams-planar}
\end{center}
\end{figure}

\paragraph{Planar vertices}. The planar vertices are the main object of interest of the present work. The term $ \frac{N}{2} \log\left(1-J^2 \hat{G} \circ \widehat{|G| \sigma}\right)$ in (\ref{action-fluctuations}) produces an infinite series of non-local (in space of two times) vertices. We will call the expression for a planar vertex without an inclusion of $\sigma$-insertions as the amputated vertex. The amputated planar vertices are constructed from the Schwinger-Dyson propagators $G$. The internal $G$-structure of an arbitrary planar vertex can be illustrated by the polygonal diagramatic technique shown on Fig.\ref{diagrams-planar}\footnote{Kitaev and Suh \cite{Kitaev17} discuss a similar diagrammatic technique in the high-temperature expansion of SYK.}. Each planar amputated vertex corresponds to a $2n$-gon, where $n$ is the order of the corresponding term in the expansion of the logarithm. The solid lines correspond to $G$, and the dashed lines indicate where the quantum field $|G|\sigma $ insertions have to be glued in in the correlation functions. Note that dashed and solid lines always interchange as one cycles through external time variables. The diagrams are shown on the Fig.\ref{diagrams-planar}. 

\section{Cutting rules for planar amplitudes}

\subsection{Derivation}

\begin{figure}[t]
\begin{center}
\includegraphics[scale=.3]{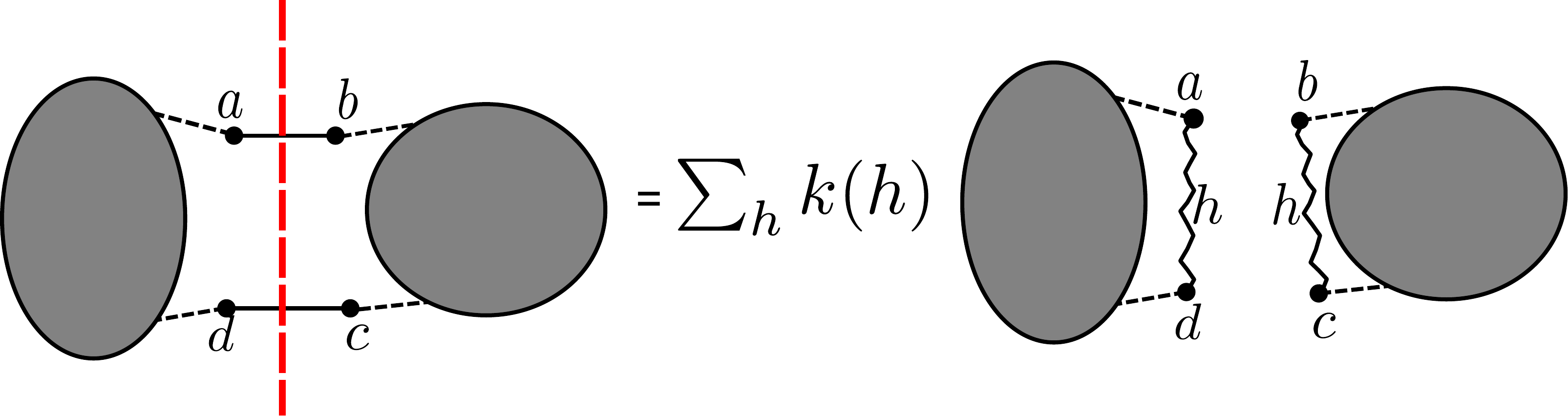}
\caption{Graphical representation of the reduction formula for planar irreducible vertices.}
\label{planar-LSZ}
\end{center}
\end{figure}

Let us consider a vertex $V$ of order $4$ or higher. Then we can always cut it in such a way that the cut crosses two solid lines and nothing else, see Fig.\ref{planar-LSZ}. Suppose that we cut through propagators $G(\tau_a, \tau_b)$ and $G(\tau_c, \tau_d)$, which connect two vertices $U_1 ({\bf \tau}_1, \tau_a, \tau_d)$ and $U_2 ({\bf \tau}_2, \tau_b, \tau_c)$. Here we denote all the time variables which are untouched by cut propagators by bold letters. In this case, we can write the vertex as 
\be
V = U_1 ({\bf \tau}_1, \tau_a, \tau_d) G(\tau_a, \tau_b) G(\tau_c, \tau_d) U_2 ({\bf \tau}_2, \tau_b, \tau_c)\,. \label{Vred}
\ee
Now we can rewrite the product of propagators through the ladder kernel using the definition (\ref{kernel}): 
\be
 G(\tau_a, \tau_b) G(\tau_c, \tau_d) = \frac{\tilde{K}(\tau_a, \tau_d; \tau_b, \tau_c)}{(q-1)J^2 | G(\tau_a, \tau_d)|| G(\tau_b, \tau_c)|}\,. \label{GG->K}
\ee
We can diagonalize the ladder kernel using the system of its (normalized) eigenfunctions $\psi_{h,n}(\tau_1, \tau_2)$: 
\be
\tilde{K}(\tau_a, \tau_d; \tau_b, \tau_c) = \sum_{h, n} k(h, n) \psi_{h, n}^* (\tau_a, \tau_d) \psi_{h, n}(\tau_b, \tau_c)\,. \label{Kdiag}
\ee
Substituting (\ref{Kdiag}) into (\ref{GG->K}) and then into (\ref{Vred}), we obtain 
\be
V = \sum_{h, n} k(h, n) U_1 ({\bf \tau}_1, \tau_a, \tau_d) \frac{\psi_{h, n}^* (\tau_a, \tau_d)}{| G(\tau_a, \tau_d)|} \frac{\psi_{h, n}(\tau_b, \tau_c)}{| G(\tau_b, \tau_c)|} U_2 ({\bf \tau}_2, \tau_b, \tau_c)\,. \label{LSZ}
\ee
Thus, we the vertex $V$ has a form of sum over products of smaller vertices $U_1$ and $U_2$, which are connected by the "partial waves" $\psi_h$. 

\subsection{Cutting symmetry identities}

\begin{figure}[t]
\begin{center}
\includegraphics[scale=.4]{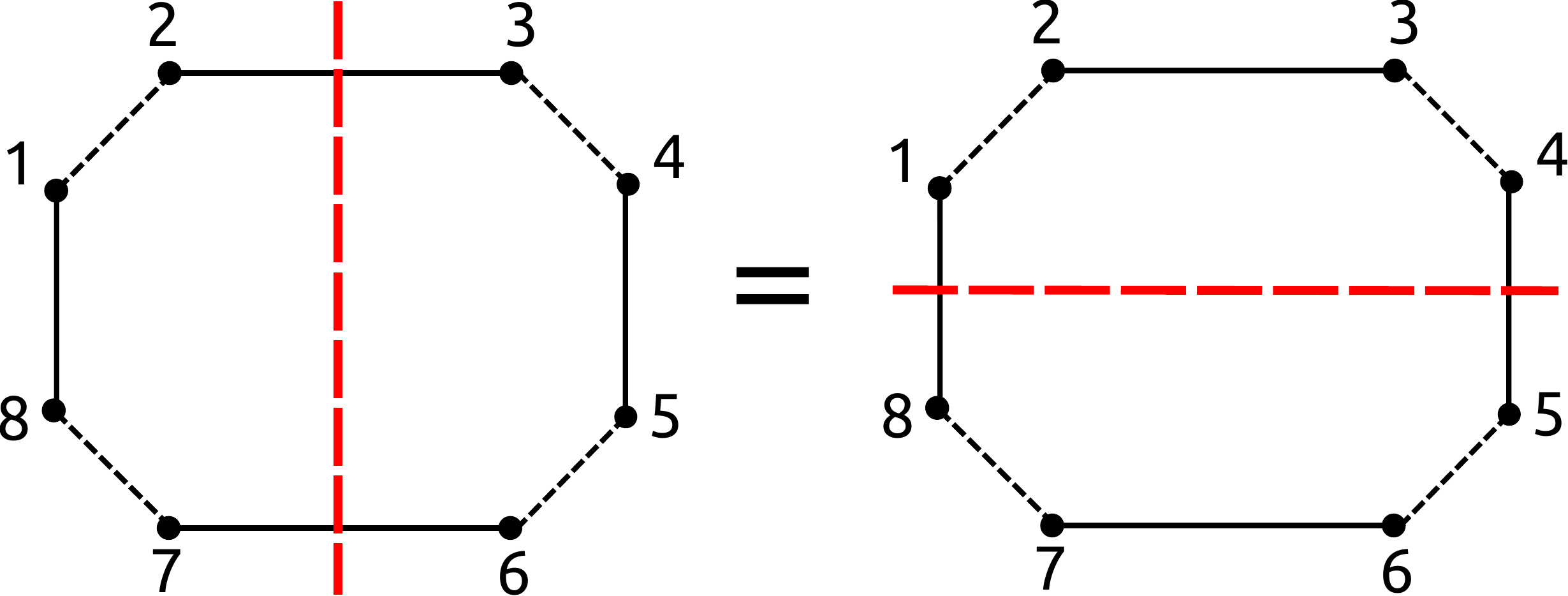}
\caption{Different cuts of the $4$-point vertex give the same result.}
\label{planar-4pt-cut}
\end{center}
\end{figure}

Let us consider the 1PI planar four-point function of the bilocals. It is obtained using the quartic vertex shown on Fig.\ref{diagrams-planar}B. The important property of the cutting rule is that it represents an equivalent way to rewrite a diagram. Therefore, different patterns of cuts on the diagrams should give the same result. In principle, one can cut the diagrams along the pairs of solid lines as much as needed, as long as each solid line is cut no more than once. The equivalence of different ways to cut gives us a set of relations reminiscent of the crossing symmetry equations. The simplest cutting symmetry equation for the $4$-point function can be obtained from the equivalence of one vertical cut and one horizontal cut, see Fig.\ref{planar-4pt-cut}. 
We can write the equation for the integrands (which in this case holds up to full derivatives): 
$$
\frac{G(t_{81})}{|G(t_{27})|}\sum_{h, n} k(h, n) \psi_{h, n}^* (t_2, t_7) \psi_{h, n}(t_3, t_6)  \frac{G(t_{45})}{|G(t_{36})|} = \frac{G(t_{23})}{|G(t_{14})|} \sum_{h, n} k(h, n) \psi_{h, n}^* (t_1, t_4) \psi_{h, n}(t_8, t_5) \frac{G(t_{67})}{|G(t_{58})|}$$

We emphasise that this relation is exact in $\beta J$, and thus can be used to constrain correlation functions beyond the conformal limit. e.g. with non-conformal contributions of the form (\ref{gg-nonconformal}) to the external lines. One can obtain similar relations for higher-point functions. 

\subsection{The six-point function example}

\begin{figure}[t]
\begin{center}
\includegraphics[scale=.4]{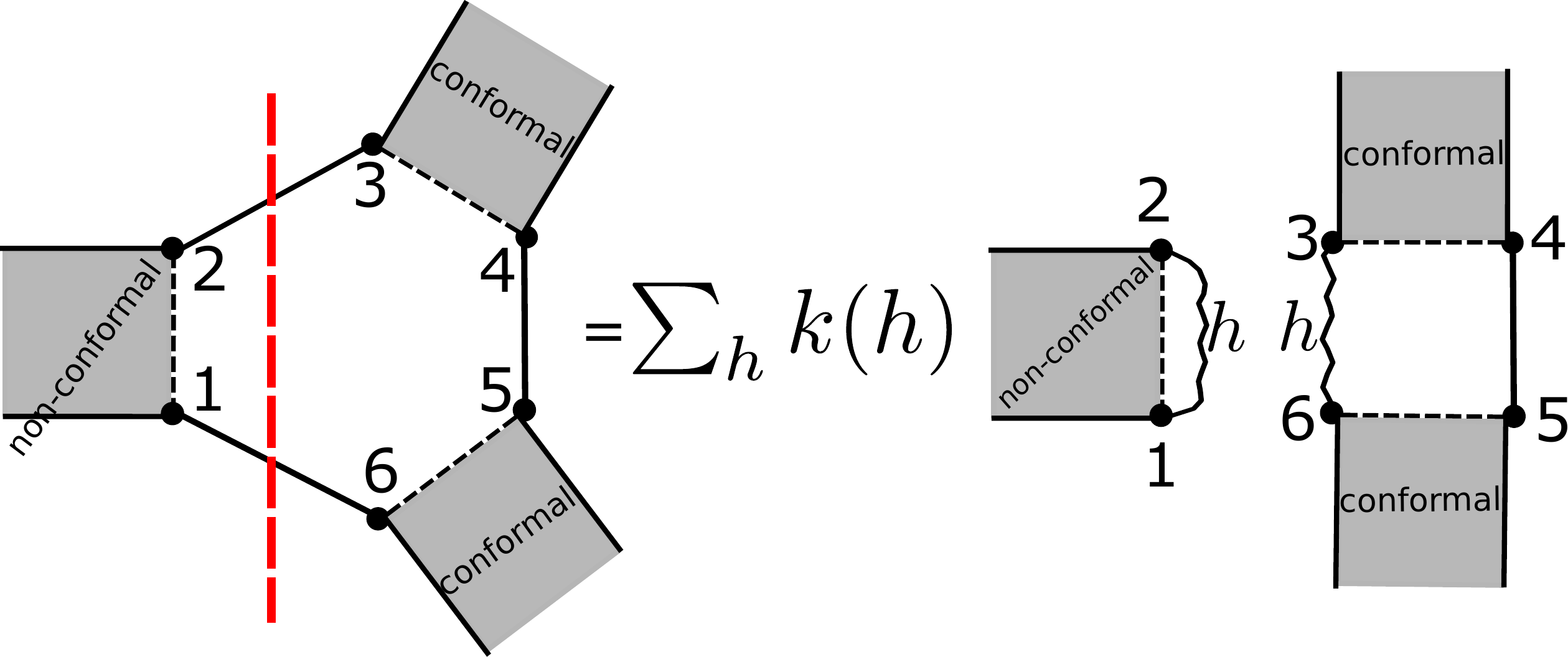}
\caption{Cutting the $3$-point function. The shaded stripes represent the external $\langle g \sigma \rangle$ propagators.}
\label{planar-3pt-cut}
\end{center}
\end{figure}

Let us consider the planar $3$-point function of bilocal fields to the leading order of $1/N$. The conformal contribution was computed in \cite{Gross17-3pt}, and it was used as one of the building blocks to construct the all point correlation functions in the conformal limit of SYK in \cite{Gross17}. In the leading $1/N$ order the planar $3$-point function is given by the $3$-point planar vertex (see Fig.\ref{diagrams-planar}A) dressed with $\langle g \sigma \rangle$ external lines: 
\be
S_2 := \langle g(\tau_1, \tau_2) g(\tau_3, \tau_4) g(\tau_5, \tau_6) \rangle_{\text{planar}} = \langle  g(\tau_1, \tau_2) g(\tau_3, \tau_4) g(\tau_5, \tau_6) V^{(3)}_\sigma \rangle\,; \label{3pt-planar-LO}
\ee
where on r.h.s. the fields are averaged with respect to the quadratic action (\ref{S2}).
We assume that one of the external lines is determined by the nonconformal contribution to the propagator, whereas the other two are conformal, as shown on the left side of the Fig.\ref{planar-3pt-cut}. The Wick contractions give a factor $3! = 6$. The vertex gives the factor of $N J^6/6$, whereas the external propagators give a factor $\frac{2}{N J^2}$ each. The expression for the planar diagram reads 
\bea
&&S_2 =  \frac{8}{N^2} \int d t_1 d t_2 d t_3 d t_4 d t_5 d t_6 \left(\frac{1}{\tK - 1}\right)(\tau_1, \tau_2; t_1, t_2) |G(t_{12})|^{\frac{q-2}{2}}\times \label{S_2}\\&& \left(\frac{1}{\tK - 1}\right)(\tau_3, \tau_4; t_3, t_4) |G(t_{34})|^{\frac{q-2}{2}} \left(\frac{1}{\tK - 1}\right)(\tau_5, \tau_6; t_5, t_6) |G(t_{56})|^{\frac{q-2}{2}} G(t_{23}) G(t_{45}) G(t_{61})\,.\nn
\eea

Our approach is to cut the vertex in such a way that the We cut the diagram through the propagators $G(t_{23})$ and $G(t_{61})$, as shown on Fig.\ref{planar-3pt-cut}. Using (\ref{LSZ}), we write the correlator as
$$
\langle g(\tau_1, \tau_2) g(\tau_3, \tau_4) g(\tau_5, \tau_6) \rangle_{\text{planar}} \equiv S_2 = 3! \frac{N J^6}{2 \cdot 3} \sum_{h, n} k(h, n) U^{(1)}_{h, n}(\tau_1, \tau_2) U^{(2)}_{h, n}(\tau_3, \tau_4; \tau_5, \tau_6)\,, $$
and we can evaluate $U$-s separately. So, 
\bea
U^{(1)}_{h, n}(\tau_1, \tau_2) = \int d t_1 d t_2 \langle g(\tau_1, \tau_2) \sigma(t_1, t_2)\rangle \frac{|G(t_{12})|}{|G(t_{12})|} \psi_{h, n}^*(t_1, t_2)\label{U1}\
\eea
 $$= \frac{2}{N J^2} \int d t_1 d t_2 \left(\frac{1}{\tK -1}\right)_{\text{non-conformal}}\!\!\!\!\!\!\!\!\!\!\!\!\!\!\!\!\!\!\!\!\!\!\!\!\!\!(\tau_1, \tau_2; t_1, t_2)  \psi_{h, n}^*(t_1, t_2) = \frac{2}{N J^2} \frac{1}{k(2, n) -1} \psi_{2, n}^*(\tau_1, \tau_2)\,. 
$$
This calculation shows that cutting away an external line of the planar diagram turns that external line simply expands the propagator in terms of the ladder eigenfunctions, and in particular, the non-conformal piece leaves just the $h=2$ eigenfunction. This reduces the number of integrations by $2$. What is left to compute is the piece
$$
U^{(2)}_{2, n}= \int d t_3 d t_4 d t_5 d t_6 \frac{|G(t_{34})||G(t_{56})|}{|G(t_{36})|} G(t_{45}) \psi_{2, n}(t_3. t_6) \langle g(\tau_3, \tau_4) \sigma(t_3, t_4)\rangle \langle g(\tau_5, \tau_6) \sigma(t_5, t_6)\rangle \,.
$$
The external lines of this piece are conformal, so one can take the OPE limit in them, analogously to the calculation of the fully conformal six-point function in \cite{Gross17-3pt}. In that case the $U^{(2)}$ reduces to a $3$-loop integral, which is much simpler to evaluate. This method of cutting off the external non-conformal lines can be applied to reduce the integral in any hugher-point planar 1PI correlation function. 

\section*{Acknowledgments}

 This contribution to the Quarks 2018 proceedings is based on the work in progress \cite{AKT} and was presented by  M.~K. at the Conference. The authors are grateful to the organizers of the Quarks 2018 conference for the opportunities to give a talk and for inspiring atmosphere during the Conference. This work is supported by the Russian Science Foundation  (project 14-50-00005, Steklov Mathematical Institute).


\begin{thebibliography}{99}
\bibitem{Sachdev92}
  S.~Sachdev and J.~Ye,
  Phys.\ Rev.\ Lett.\ 70, 3339 (1993)

\bibitem{Kitaev} A.~Kitaev, talks at KITP in 2015:\\ http://online.kitp.ucsb.edu/online/entangled15/kitaev/,\\ http://online.kitp.ucsb.edu/online/entangled15/kitaev2/

\bibitem{MScomments} 
  J.~Maldacena and D.~Stanford,
  Phys.\ Rev.\ D {\bf 94}, no. 10, 106002 (2016)
  [arXiv:1604.07818 [hep-th]].

\bibitem{Kitaev17} 
  A.~Kitaev and S.~J.~Suh,
  JHEP {\bf 1805}, 183 (2018)
  [arXiv:1711.08467 [hep-th]].

\bibitem{Sachdev15} 
  S.~Sachdev,
  Phys.\ Rev.\ X {\bf 5}, no. 4, 041025 (2015)
  [arXiv:1506.05111 [hep-th]].

\bibitem{Maldacena16} 
  J.~Maldacena, D.~Stanford and Z.~Yang,
  PTEP {\bf 2016}, no. 12, 12C104 (2016)
  [arXiv:1606.01857 [hep-th]].

\bibitem{Gross17-3pt} 
  D.~J.~Gross and V.~Rosenhaus,
  JHEP {\bf 1705}, 092 (2017)
  [arXiv:1702.08016 [hep-th]].

\bibitem{Gross17} 
  D.~J.~Gross and V.~Rosenhaus,
  JHEP {\bf 1712}, 148 (2017)
  [arXiv:1710.08113 [hep-th]].

\bibitem{Bagrets16} 
  D.~Bagrets, A.~Altland and A.~Kamenev,
  Nucl.\ Phys.\ B {\bf 911}, 191 (2016)
  [arXiv:1607.00694 [cond-mat.str-el]].

\bibitem{Mertens17} 
  T.~G.~Mertens, G.~J.~Turiaci and H.~L.~Verlinde,
  JHEP {\bf 1708}, 136 (2017)
  [arXiv:1705.08408 [hep-th]].

\bibitem{Polchinski16} 
  J.~Polchinski and V.~Rosenhaus,
  JHEP {\bf 1604}, 001 (2016)
  [arXiv:1601.06768 [hep-th]].

\bibitem{AKT}
I.~Ya.~Aref'eva, M.~A.~Khramtsov, M.~D.~Tikhanovskaya, to appear

\bibitem{AKTV}
I.~Aref'eva, M.~Khramtsov, M.~Tikhanovskaya and I.~Volovich, ``Replica-nondiagonal solutions in the SYK
model,'' in preparation. 

\bibitem{AKTV-Quarks}
I.~Aref'eva, M.~Khramtsov, M.~Tikhanovskaya and I.~Volovich, ``On replica-nondiagonal large $N$ saddles in the SYK model,'' Quarks 2018 proceedings.

\bibitem{AV}
 I.~Aref 'eva and I.~Volovich,
  ``Notes on the SYK model in real time,'' Theoret. and Math. Phys., vol.196 (2018),
  arXiv:1801.08118 [hep-th].
  
\end{thebibliography}
\end{document}